\documentclass[pre,twocolumn,showpacs,floatfix,amsmath,amssymb]{revtex4}

\usepackage{graphicx}
\usepackage{dcolumn}
\usepackage{color}
\usepackage{bm}
\usepackage{verbatim}
\usepackage{multirow}

\newcommand{\tableOne}{
\begin{table}[b]
	\begin{ruledtabular}
	\begin{tabular}{c c l l l l }
		Model & Evolution & $Q_{1}^{*}$ & $Q_{2}^{*}$ & $Q_{3}^{*}$ & $Q_{4}^{*}$\\
   		\hline \hline
    	\multirow{2}{*}{SP} & EW & 0.18 & 0.09 & 0.07 & 0.06 \\
        		& KPZ & 0.16 & 0.08 & 0.06 & 0.04 \\
        \hline
    	\multirow{1}{*}{CD} & EW/KPZ & 0.22 & 0.11 & 0.08 & 0.055 \\
	\end{tabular}
	\end{ruledtabular}
	\caption{ $Q_{m}^{*}$ values for SP and CD models.}
  	\label{tab:1}
\end{table}
}

\newcommand{\figOne}{
	\begin{figure*}[t]
		\begin{center}
			\includegraphics[width=0.95\textwidth]{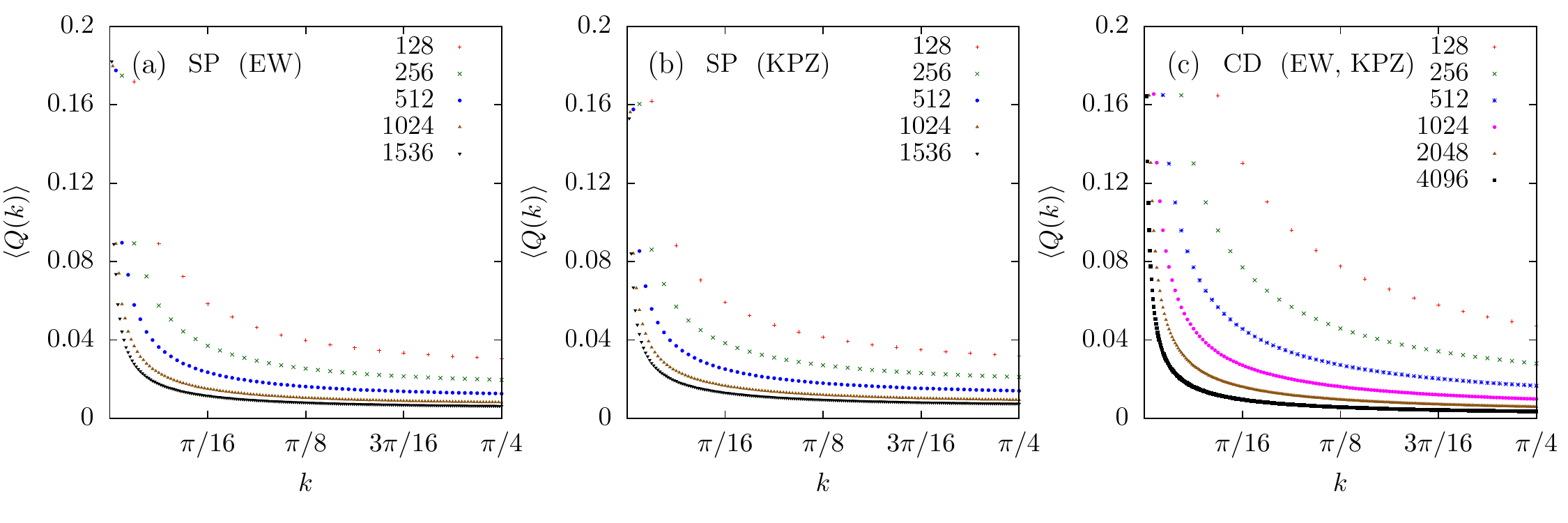}
		\end{center}
	
		\caption{ (Color online) The plots of $\langle Q(k) \rangle$ versus $k$ for different
			system sizes for (a) SP model with EW, (b) SP model with KPZ, and (c) CD model with
			EW/KPZ surface evolutions suggests that $\langle Q(k) \rangle$ approaches a finite
			value as $k={2\pi m}/{L} \to 0$ with $m$ held fixed. The limiting value for each
		$m$ provides a partial glimpse of the ordering in the system. \label{fig:QkkCD}}

	\end{figure*}
}

\newcommand{\figTwo}{
\begin{figure*}[htbp]
	\begin{center}
    \includegraphics[width=0.75\textwidth]{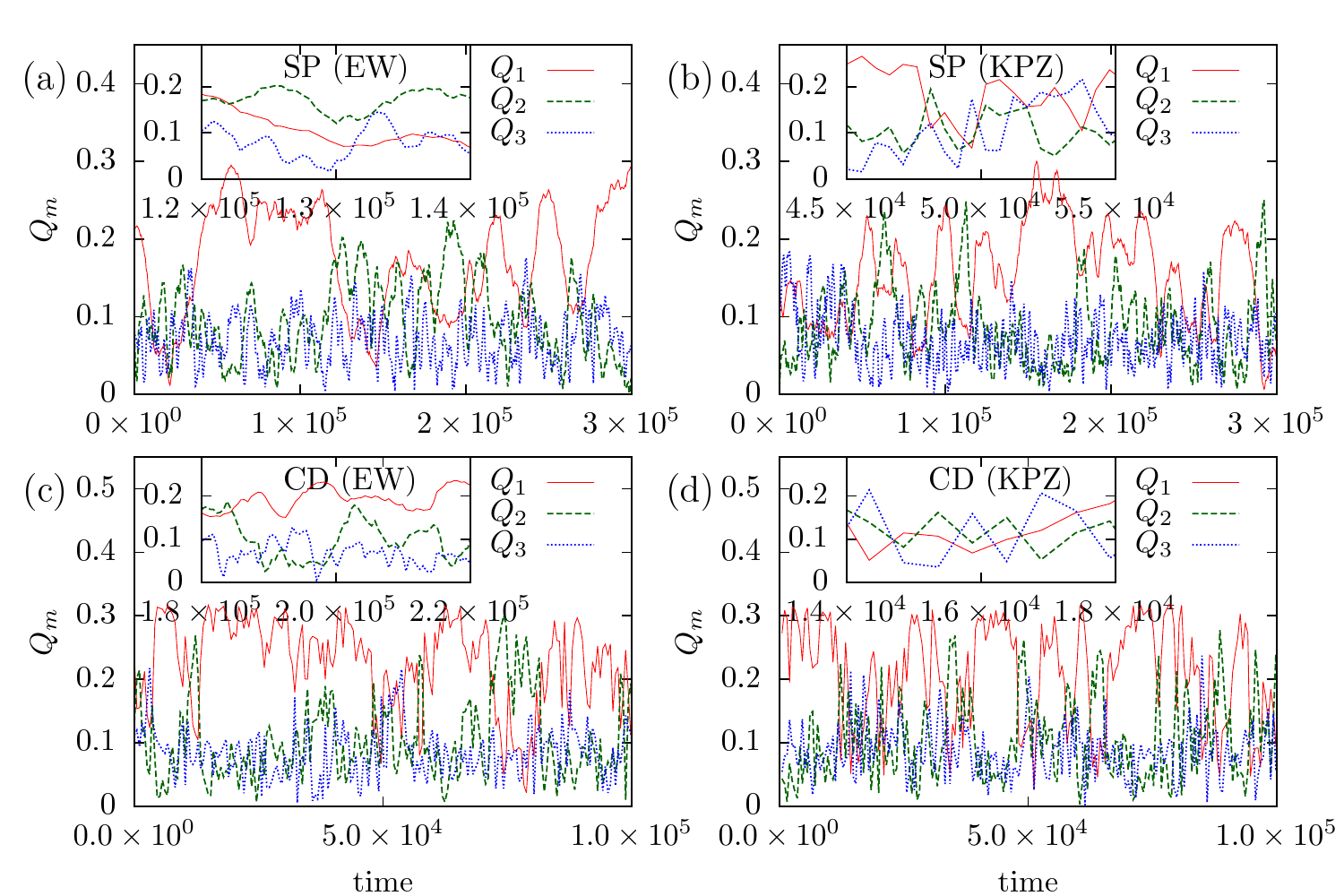}
  \end{center}

	\caption{ \label{fig:qtime} (Color online) The variation of first three modes of the order
		parameter $Q_{m}$ as a function of time with a system size $L=512$ are plotted for (a)
		SP model with EW surface evolution, (b) SP model with KPZ surface evolution, (c) CD
		model with EW surface evolution, and (d) CD model with KPZ surface evolution. These time
		series of $Q_{m}$ show that an increase in value of one mode is usually accompanied by the
		decrease in the others and vice versa.
	}

\end{figure*}
}

\newcommand{\figThree}{
\begin{figure*}[htpb]
 	\begin{center}
    \includegraphics[width=0.95\textwidth]{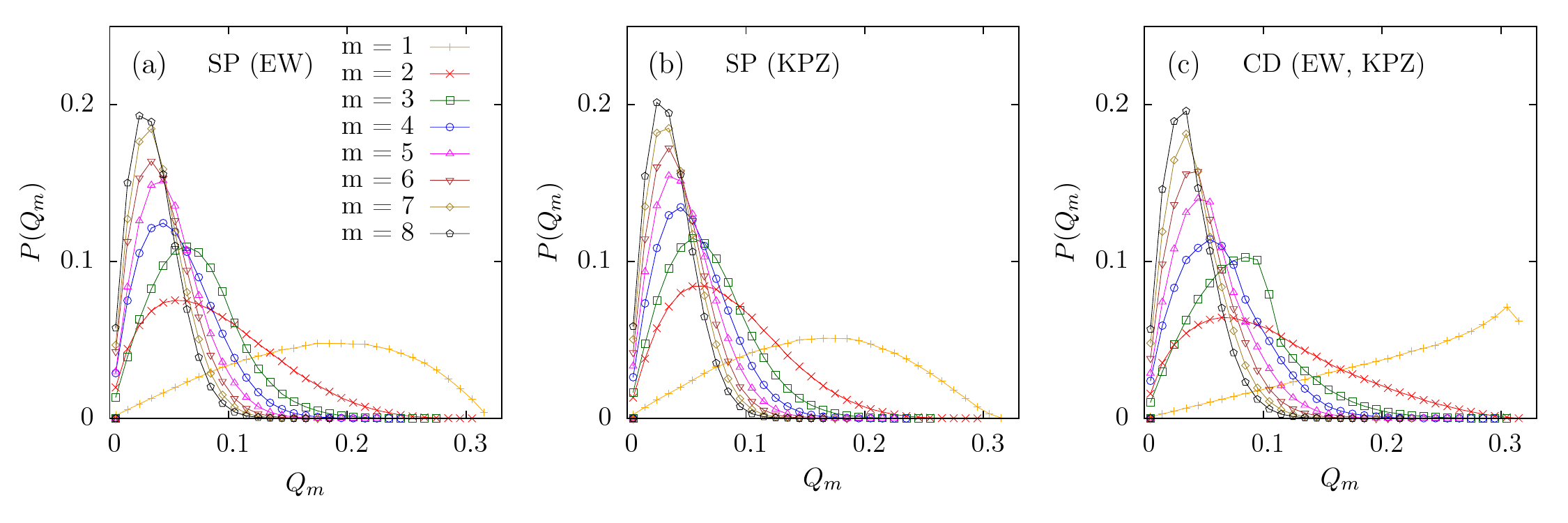}
  \end{center}
 
	\caption{(Color online) Probability distribution function $P(Q_{m})$ for different modes of
		the order parameter $Q_m$ ($m$ is the mode number) are plotted for (a) SP model with EW
		surface evolution for lattice size $L=192$, (b) SP model with KPZ surface evolution for
		lattice size $L=512$ and (c) CD model with EW/KPZ surface evolution for lattice size
		$L=192$.  
	} \label{fig:probability}

\end{figure*}
}

\newcommand{\figFour}{
\begin{figure*}[htpb]
  	\begin{center}
		\includegraphics[width=0.95\textwidth]{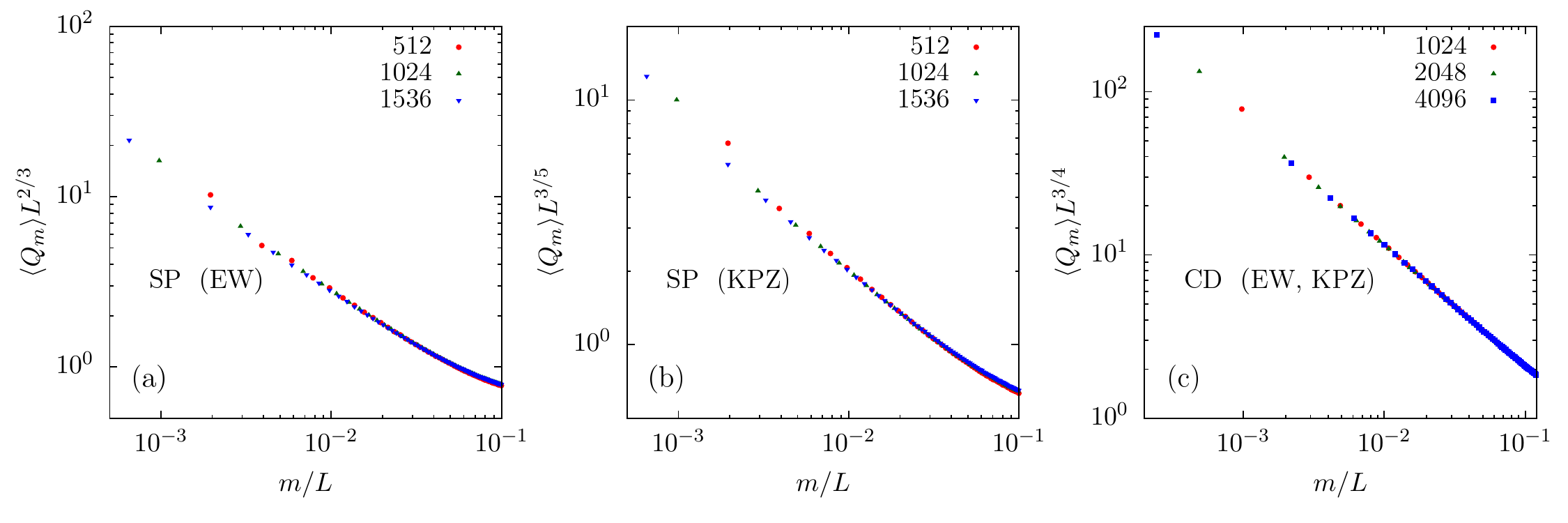}
	\end{center}

	\caption{(Color online) The plot of one-point function or order parameter $\langle Q_m
		\rangle $ as a function of mode number $m$ when it is scaled by $L^{\phi}$ suggests
		that the exponent $\phi$ for (a) SP model with EW surface evolution is close to $2/3$ ,
		(b) SP model with KPZ surface evolution is close to $3/5$, and (c) CD model with EW/KPZ
		surface evolution is close to $3/4$.
	} \label{fig:scaling}

\end{figure*}
}

\newcommand{\figFive}{
\begin{figure*}[htpb]
  \begin{center}
	  \includegraphics[width=0.95\textwidth]{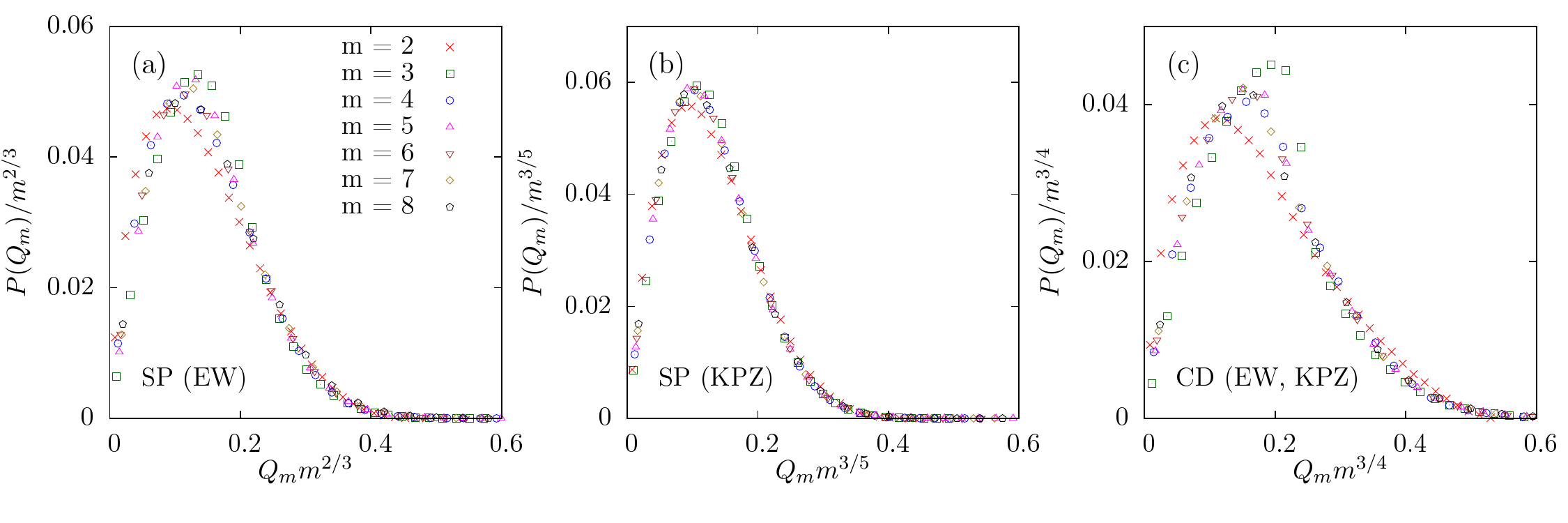}
  \end{center}

  \caption{ (Color online) The plot of scaled probability distribution functions
	  $P(Q_m)/m^{\phi}$ for different modes  of the order parameter, as a function of order
	  parameter $Q_m$ when it is scaled by $m^{\phi}$ with values of $\phi$ taken from
	  Fig.~\ref{fig:scaling} for $\langle Q_m \rangle$, we find the curves collapse for larger
	  values of $m$. } \label{fig:probscaling}

\end{figure*}
}

\newcommand{\figSix}{
\begin{figure}[t]

	\begin{center}
		\includegraphics[width=0.48\textwidth]{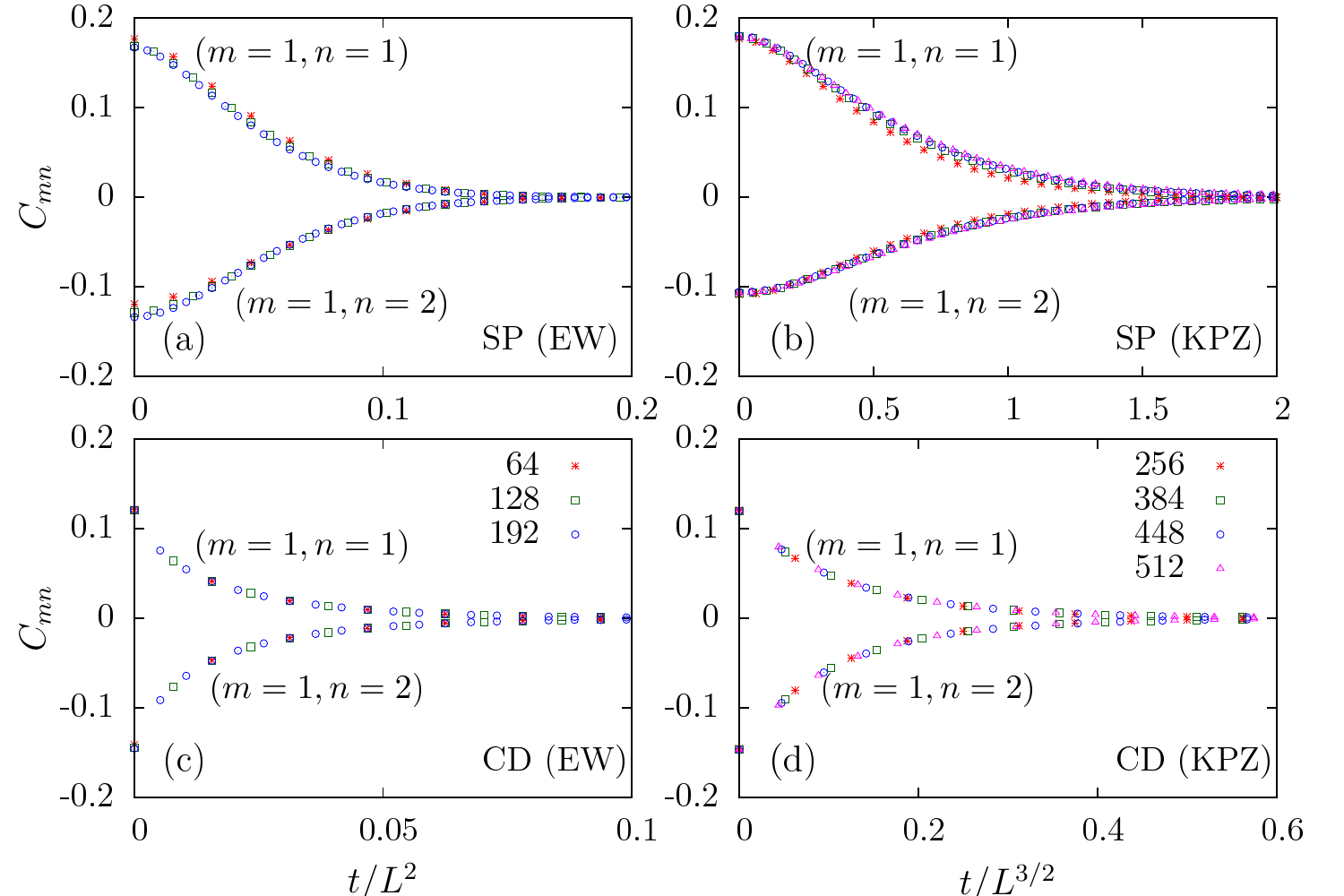}
	\end{center}

	\caption{(Color online) The plot of scaled correlation function of order parameter for both
		auto-correlation $\langle Q_m(0)Q_m(t) \rangle$ and cross-correlation $\langle Q_m(0)Q_n(t)
		\rangle$  for (a) SP model with EW, (b) SP model with KPZ, (c) CD model with EW, (d) CD
		model with KPZ surface evolutions show that steady state auto-correlations decay with
		time and scale with $t/L^z$ where $L$ is the system size and $z$ is the dynamic exponent.
		However, cross-correlations of order parameter increase with time from their initial
		negative values implying anticorrelations decay in times. } \label{fig:corrscale}

\end{figure}
}

\newcommand{\figSeven}{
\begin{figure*}[htbp]
	\begin{center}
  		\includegraphics[width=\textwidth]{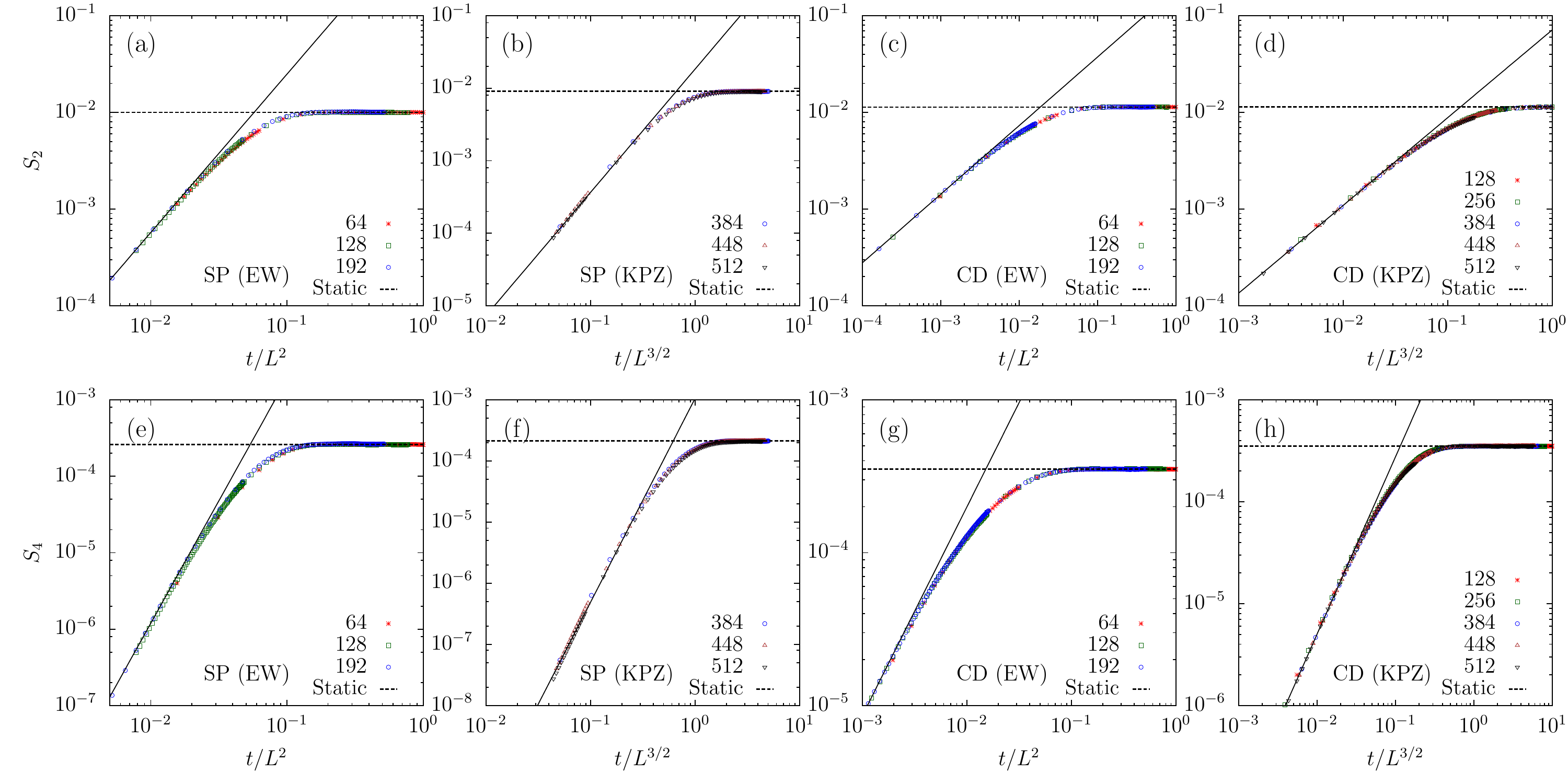}
	\end{center}

	\caption{(Color online) Structure factors $S_2(y)$ and $S_4(y)$ for different system sizes
		as a function of the scaled time $y = t/L^z$ for SP and CD models with both EW and KPZ
		surface evolutions. The saturation values of $S_2$ and $S_4$ are determined from steady
		state data and is shown by dashed lines. The solid lines indicate the slope at small
		values of $y$. } \label{fig:S2S4}
		
\end{figure*}
}

\newcommand{\figEight}{
\begin{figure}[b]
	\begin{center}
  		\includegraphics[width=0.48\textwidth]{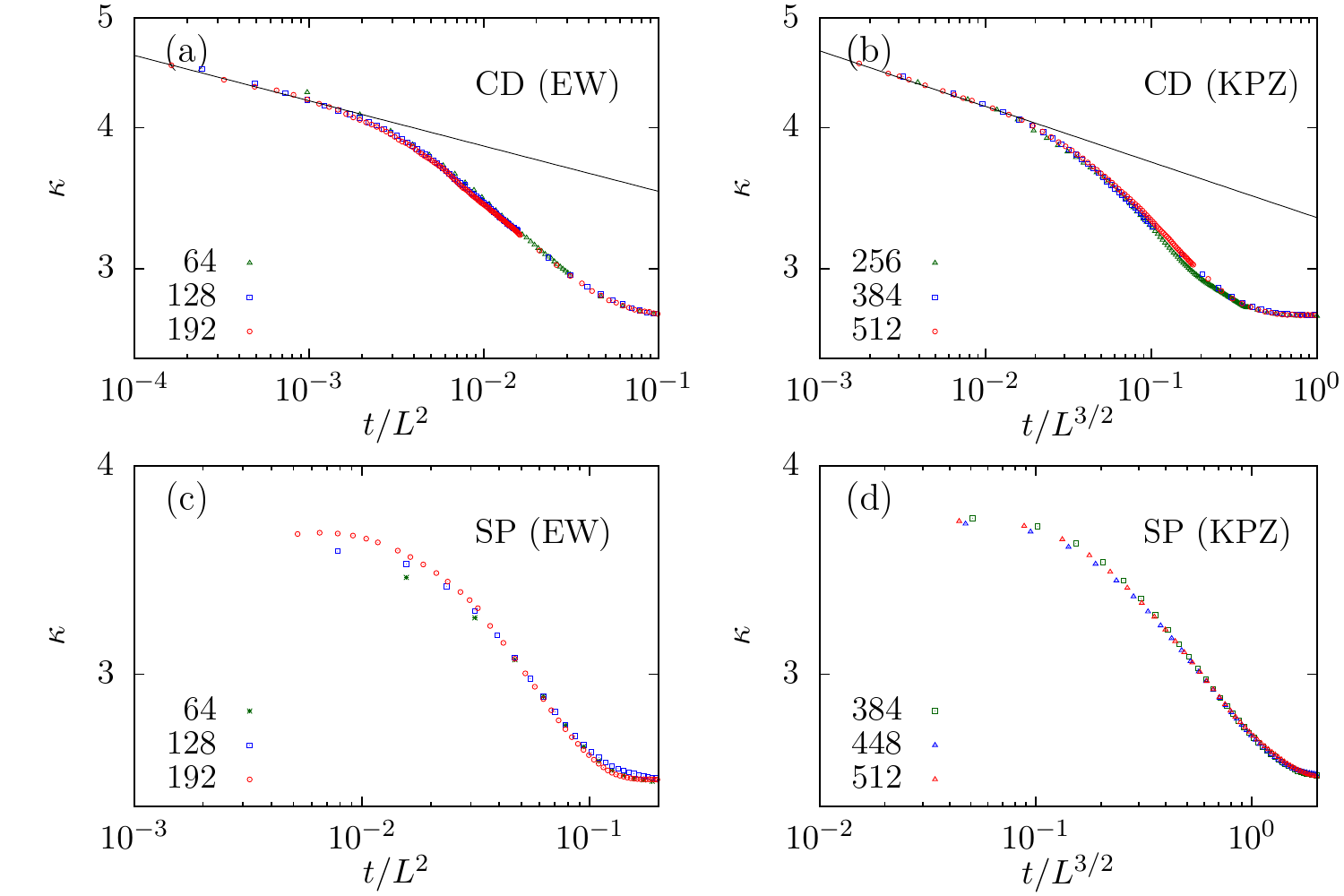}
	\end{center}

	\caption{(Color online) The flatness, $\kappa$, of the first Fourier mode is plotted as a
		function of $t/L^{z}$ for various lengths. The weak divergence of the flatness seen for
		the CD model indicates intermittency. } \label{fig:flat}

\end{figure}
}

\begin{document}

\title {Order parameter scaling in fluctuation dominated phase ordering}

\author{Rajeev Kapri}
\email{rkapri@iisermohali.ac.in}
\affiliation{Department of Physical Sciences, Indian Institute of Science Education and
Research Mohali, Sector 81, Knowledge City, S. A. S. Nagar, Manauli PO 140306, India. }

\author{Malay Bandyopadhyay}
\email{malay@iitbbs.ac.in}
\affiliation{School of Basic Sciences, Indian Institute of Technology Bhubabneswar, Satyanagar,
Bhubaneswar 751007, India.}

\author{Mustansir Barma}
\email{barma@theory.tifr.res.in}
\affiliation{Department of Theoretical Physics, Tata Institute of
Fundamental Research, Homi Bhabha Road, Colaba, Mumbai 400005, India.}

\date{\today}

\begin{abstract}

	In systems exhibiting fluctuation-dominated phase ordering, a single order parameter does
	not suffice to characterize the order, and it is necessary to monitor a larger set. For
	hard-core sliding particles (SP) on a fluctuating surface and the related coarse-grained
	depth (CD) models, this set comprises the long-wavelength Fourier components of the density
	profile. We study both static and dynamic scaling laws obeyed by the Fourier modes $Q_m$
	and find that the mean value obeys the static scaling law $\langle Q_m \rangle \sim
	L^{-\phi}f(m/L)$ with $\phi\simeq2/3$ and $\phi\simeq3/5$ with Edwards-Wilkinson (EW) and
	Kardar-Parisi-Zhang (KPZ) surface evolution respectively. The full probability distribution
	$P(Q_m)$ exhibits scaling as well. Further, time-dependent correlation functions such as
	the steady state auto-correlation and cross-correlations of order parameter components are
	scaling functions of $t/L^z$, where $L$ is the system size and $z$ is the dynamic exponent
	with $z=2$ for EW and $z=3/2$ for KPZ surface evolution. In addition we find that the CD
	model shows temporal intermittency, manifested in the dynamical structure functions of the
	density and a weak divergence of the flatness as the scaled time approaches zero.

\end{abstract}
\maketitle

\section{Introduction}

In equilibrium statistical mechanics, the notion of an ordered state is well understood and the
degree of order is characterized with the aid of two-point and one-point functions. Thus, the
long-range order (LRO) $m_c^2$ is defined through the asymptotic behaviour of a two-point
correlation function, while the order parameter $m_s$, the spontaneous value in a vanishing
field, is a one-point function. In the thermodynamic limit, fluctuations become insignificant
and $m_s$ and $m_c$ have well defined values \cite{griffith1}. Further, for systems with
short-ranged interactions, conditions under which ordering is possible are known. Ordering
occurs only if the spatial dimension exceeds one for a scalar order parameter, or exceeds two
if ordering breaks a continuous symmetry \cite{griffith2}.

In nonequilibrium steady states, other types of behaviour are possible and characterization of
order needs to be addressed afresh. Of particular interest is a class of systems in which
fluctuations are anomalously strong, but which nevertheless have a propensity to order, leading
to fluctuation-dominated phase ordering (FDPO)\cite{mustansir1,mustansir2,manoj}. FDPO arises
in several types of systems. These include passive scalar systems, in which a passive species
is driven by an autonomously evolving field, e.g., particles driven by a fluctuating surface or
a noisy Burgers fluid in one and two dimensions \cite{mustansir1,mustansir2,manoj}, or, in the
context of active systems, by a 2D nematic field \cite{mishra}. FDPO also arises in a model 1D
granular gas in which the coefficient of restitution depends on the velocity of approach
\cite{shinde}. The signature of FDPO is a cusp singularity at small argument in the scaled
two-point correlation function, signifying a breakdown of the Porod law. The relation of this
singularity to giant number fluctuations has been discussed in Ref.~\cite{dey} in the context
of active systems.  Stronger singularities of the scaling function arise if the passive
particles do not have hard core interactions \cite{nagar}. Finally, there seems to be an
intriguing connection between FDPO and systems with quenched disorder, and cusp singularities
have been observed in a number of systems with rough surfaces and
interfaces~\cite{bale,wong,sorensen,shrivastav1,shrivastav2}.

This paper deals with the characterization of order in a passive scalar system which exhibits
FDPO. The system consists of hard-core particles sliding locally downwards, along the local
slopes of a one-dimensional fluctuating surface with overall slope
zero~\cite{mustansir1,mustansir2}. The surface evolves through its own dynamics whereas the
particle movement is guided by local surface gradients. As time passes, particles are driven
towards each other and the spatial extent of particle-rich regions increases as a function of
time, until it is of the order of system size in steady state. 

In earlier work, this tendency  was quantified by studying coarsening properties, as the system
evolves in time starting from a completely disordered state. As usual for coarsening systems,
the equal time correlation function $C(r,t)$ obeys scaling, with a growing length scale
${\mathcal{L}}(t) \sim t^{1/z}$.  The unusual point here is that the scaling function shows a
cusp singularity as the argument $r/\mathcal{L}(t)$ approaches zero. In momentum space, this
translates into the scaled structure factor varying as $\lbrack k \mathcal{L}(t) \rbrack
^{-(1+\alpha)}$ where $k$ is the wave vector. The exponent $\alpha$ is less than 1 signifying a
cusp singularity, which represents a marked deviation from the linear behaviour characteristic
of the Porod law ($\alpha=1$)  which holds normally for scalar field coarsening
\cite{porod,bray}.  Further, cusp singularities were also observed in the correlation function
in steady state \cite{mustansir1,mustansir2}, and also in the decay of the temporal
auto-correlation function~\cite{sakuntala}.

In this paper, our focus is on the study of both static and dynamic aspects of the order
parameter or one point function $Q_m(k,t)$. We argue that a single scalar order parameter does
not suffice to properly characterize the order and we need to monitor a larger set. This set is
built from the long-wavelength Fourier components of the density profile.  We present evidence
that these Fourier components have  probability distributions which remain broad in the
thermodynamic limit; that they are anticorrelated with each other; and that the probability
distributions of the components for different sizes are described by simple scaling laws. We
supplement our studies of the sliding particle system by studying a related coarse-grained
depth model, and show that this model exhibits broadly similar behaviour. Further, the temporal
behaviours (of autocorrelation and cross-correlation of the Fourier modes) are studied and
scaling properties are investigated. This includes a study of dynamical structure function and
the flatness, which is found to exhibit a divergence at small argument, indicating that the
behaviour in time is intermittent~\cite{frisch,sachdeva}.

The paper is organized as follows: In Sec.~\ref{sec:model}, we introduce the sliding particle
(SP) and coarse-grained depth (CD) models of fluctuating surfaces and review the behaviour of
the two-point correlation function. In Sec.~\ref{sec:statics}, we discuss the static properties
of one point function, arguing that long-wavelength Fourier transforms of the density profile
constitute an appropriate set of order parameters. We discuss their probability distributions,
the correlations between them, and  scaling properties. In  Sec.~\ref{sec:dynamics}, we discuss
the corresponding dynamical properties including cross correlations between modes. We conclude
with a discussion of our results in Sec.~\ref{sec:conclude}.

\section{The Model and Two-Point Correlation Functions} \label{sec:model}

In this section, we discuss some simple models that show FDPO. The sliding particle (SP) model
involves a system of hard core particles sliding under gravity on a stochastically evolving
surface \cite{mustansir1,mustansir2,manoj} through either Edwards-Wilkinson (EW) or
Kardar-Parisi-Zhang (KPZ) dynamics \cite{ew,kpz}. We also define the coarse grained depth (CD)
model which only involves the height field of the fluctuating surface, and which provides
considerable insight. We then summarize the results of \cite{mustansir1,mustansir2} on the
scaling behaviour of two-point correlation functions in these models.

\subsection{Sliding particle (SP) model}

This is a lattice model of particles moving on a 1D fluctuating
surface~\cite{mustansir1,mustansir2}. The particles and links are represented respectively by
$\lbrace \sigma_i\rbrace$ and $\lbrace \tau_{(i-1/2)}\rbrace$.  Both $\sigma$'s and $\tau$'s
are Ising like variables that can take values $\pm 1$ on a one dimensional lattice with
periodic boundary conditions. The occupation of any site $i$ can be given in terms of
$\sigma_i$'s, as $n_i=\frac{1}{2}(1+\sigma_i)$ ($n_i = 0, 1$), and the local slope of the
surface connecting site $i$ is given by the variable $\tau_{(i-1/2)}=+1$ or $-1$ and is denoted
by / or $\backslash$ respectively.

The evolution of the surface can be modelled by either the EW or the KPZ dynamics. This is
incorporated in our model by the stochastic corner flips involving exchange of adjacent
$\tau$'s: thus $ \wedge \rightarrow \vee $ occurs with rate $p_1$, while $\vee \rightarrow
\wedge$ changes with rate $q_1$. The symmetric surface fluctuations, i.e. $p_1 = q_1$, belong
to the EW class whereas $p_1 \ne q_1$ belongs to the KPZ class.  The dynamics of the particles
can be modelled by exchanging the position of a particle and a vacancy (hole) pair at adjacent
sites $(i, i+1)$ with rates that depend on the intermediate local slope: thus the moves
$\bullet \backslash \circ \rightarrow \circ \backslash \bullet$ and $\circ / \bullet
\rightarrow \bullet / \circ$ occur at a rate $p_2$, while the reverse moves occur at a rate
$q_2 < p_2$. This asymmetry reflects the fact that it is easier to move downwards under the
influence of a gravitational field. In our study we consider the strong field limit i.e., $q_2=0$
for the particle system and set $p_2 = p_1$.  The  dynamics conserves both $\sum \sigma$ and
$\sum \tau$, and we study that sector where $\sum\sigma$ and $\sum\tau$ vanish corresponding to
the half-filled system of particles on a surface with zero average tilt.

\figOne

\subsection{Coarse-grained depth (CD) model}

In order to describe the dynamics of the hills and valleys of the surface, let us consider the
height profile $\lbrace h_i \rbrace$ with $h_i =\sum_{ 1 \le j \le i} \tau_{j-1/2}$.  We wish
to define a model involving only $\lbrace h_i\rbrace$, to mimic the SP model. Since SP
particles slide down and occupy the lower portion of the height profile, we define an analogous
variable $\sigma_i$ in the CD model as taking on the values $+1$, $-1$, or $0$ depending on
whether the surface profile at site $i$ is below, above, or exactly coincident with a reference
level: $\sigma_i = - sgn \lbrack h_i - \langle h(t) \rangle \rbrack$, where $\langle h(t)
\rangle = \frac{1}{L} \sum_{i=1}^L h_i(t) $ is the instantaneous average height which
fluctuates with time. The nomenclature Coarse-grained Depth model derives from the fact that
the mapping $h_i\rightarrow \sigma_i$ may be viewed as a coarse graining which eliminates all
fluctuations of the height other than the sign. The CD model is concerned with the static and
dynamic properties of $\lbrace \sigma_i\rbrace$, which mimics the particle configuration in the
SP model.

\subsection{Two-point correlation function : overview}

The two-point correlation function characterizes the steady state approached by the system of
particles driven by a fluctuating surface that exhibits LRO. We monitor the evolution of the
local density using Monte Carlo simulations. For both the EW and KPZ surfaces, no equilibration
is needed because every configuration of hills and valleys with periodic boundary condition
carries equal weight and can be chosen at random as a valid surface configuration in the steady
state.  The particles are distributed randomly on sites. In every Monte Carlo step, we
performed $2L$ updates ($L$ each for sites and bond variables) at random.  The density
distribution is guided by the evolution of the surface profile.

To quantify the tendency towards clustering, one can define the equal time two point
correlation function as $C(r,t) = \langle \sigma_i(t) \sigma_{i+r}(t) \rangle$ for the SP
model, and $C(r,t) = \langle s_i(t) s_{i+r}(t) \rangle$ for the CD model. The correlation
functions have the scaling form
\begin{equation}
	C(r,t) = Y\Big(\frac{r}{{\mathcal{L}}(t)}\Big),
\end{equation}
where ${\mathcal{L}}(t)$ is a time-dependent length scale which describes the typical linear size
of a cluster at time $t$, typically growing as a power law in time i.e. ${\mathcal{L}}(t)\sim
t^{1/z}$. Here, $z$ is the dynamical exponent characteristic of the surface fluctuations. This
length scale ${\mathcal{L}}(t)$ for density fluctuations is  set by the base lengths of typical
coarse-grained hills which have overturned in time $t$.

In 1D, we have $z=2$ for EW and $z=3/2$ for KPZ surface evolution. The existence of such a
single growing length scale is a signature of coarsening towards a phase-ordered state. The
coarsening is driven by surface fluctuations, rather than temperature quenching. As previously
studied in Refs.~\cite{mustansir1,mustansir2} the scaling curves have cusps at small values of
$r/{\mathcal{L}(t)}$ with a nonzero intercept. That is, $ C(y) = C_0 - C_1 y^{\alpha}$. The
intercept $C_0$ is equal to the LRO measure $m_0^2 = \lim_{r\rightarrow\infty} \langle
\sigma_i\sigma_{i+r} \rangle$. This is because an arbitrary large but fixed distance $r$
corresponds to $y={r}/{\mathcal {L}(t)} \rightarrow 0$  in a coarsening system, since $\mathcal
{L}(t)$ diverges as $t\rightarrow \infty$. The cusp exponent $\alpha$ is found numerically to
be $\simeq 0.5$ for EW evolution and $\simeq 0.25$ with KPZ dynamics of the
interface~\cite{mustansir1,mustansir2}. In fact, the cusp exponent $\alpha=1/2$ can be found
analytically for the CD model. The cusp implies a tail of scaled structure factor $S(k) \sim
(k{\mathcal{L}})^{-(1+\alpha)}$ for large $k{\mathcal{L}}$. This shows a deviation from the
Porod law $S(k) \sim (k{\mathcal{L}})^{-2}$, characteristic of customary coarsening.

If instead of considering coarsening in an infinite system, we consider steady state in a
finite system of size $L$, we may replace $\mathcal{L}(t)$ by $L$ to conclude that the scaled
two-point correlation function in the steady state can be written as
\begin{equation}
  C_s \left( \frac{r}{L} \right) = m_0^2 - c_1 \left| \frac{r}{L} \right|^{\alpha},
\end{equation}
with $\mid r/L\mid << 1$. Here, as with coarsening, the cusp exponent $\alpha = 0.5$ for EW
and $\alpha = 0.25$ for KPZ surface evolution.

\figTwo

\section{Order Parameter Statics} \label{sec:statics}

The long-wavelength Fourier transforms of the density profile (in the SP model) or the
coarse-grained depth (in the CD model) constitute a set of order parameters appropriate to
characterize the type of large-scale clustering that sets in. In this section, we study the
probability distributions of these parameters and their scaling properties as the system size
is changed.

\subsection{The order parameter set}

Let us define Fourier components of the density
\begin{equation}
  Q(k)=  \left | \frac{1}{L}\sum_{j=1}^L  n_j \exp \left(i \frac{2\pi m j}{L} \right) \right |,
\end{equation}
where, $n_j = \frac{1}{2}(1+\sigma_j)$ and $m=1,2,\cdots L-1$. As we will see below, the lowest
nonzero Fourier components $Q_{m}^{*}= \langle Q \left( {2\pi m}/{L} \right) \rangle$ can
be used as a measure of the phase separation in our system with conserved dynamics.

Figure \ref{fig:QkkCD} shows the mean values $\langle Q(k) \rangle$ as a function of $k$ for
various values of $L$ for SP and CD models. Two points should be noted. (a) First, for any
fixed non zero value of wave-vector $k$, one observes that $\langle Q(k) \rangle \rightarrow 0$
as $L \rightarrow \infty$. (b) In order to study the $k \rightarrow 0$ limit, it is instructive
to observe the behaviour of the low-$m$ modes $Q_m = Q \left( {2\pi m}/{L} \right)$ with
$m=1,2,3,4, \ldots$ and monitor $Q_m^{*} \equiv \langle Q_m \rangle = \langle Q \left(
{2m\pi}/{L} \right) \rangle$. It is seen from Fig. \ref{fig:QkkCD} that $Q_1^*$, $Q_2^*$,
etc., each approach a finite limit as $L\rightarrow \infty$.

The values of first four Fourier modes for various models are tabulated in Table~\ref{tab:1}
for both the EW and KPZ surface evolutions. From the table, we observe that $Q_{1}^{*}$ is the
largest, and it thus provides a first characterization of order.

\tableOne

The values of $Q_1^*, Q_2^*,Q_3^*,\ldots$ characterize the gross form of the macroscopic
cluster that occurs in the system. For example, $Q_1^*$ is largest in configurations with a
single dense cluster of particles, $Q_2^*$ is largest in configurations with two well separated
dense clusters and so on.  The mean values of the Fourier components $Q_m^*$ decrease with
increasing $m$, providing strong evidence that macroscopic clustering occurs in the system.

Figure~\ref{fig:qtime} shows the time series for the long-wavelength Fourier modes $Q_m$.  The
strong excursions about their average values lead to broad probability distributions. In fact
these distributions remain broad in the thermodynamic limit. This is contrary to the situation
in normal equilibrium phase transitions where the order parameter has a well defined value in
the thermodynamic limit and its distribution consists of delta functions at values that
characterize the phase.

\figThree

In Fig.~\ref{fig:probability}, we have shown the probability distribution functions of the
first eight modes $Q_m \ (m=1,2,\ldots,8)$ for both the SP and CD models with EW and KPZ
surface evolutions. These distributions are obtained numerically for system sizes $L=64$, 128,
and 192 for EW, and $L=128$, 256, and 512 for KPZ surface evolutions. It is evident from the
figure that the distributions for various modes are spread over a wide range of  values.
Furthermore, these distributions approach a limiting form on increasing the system size; they
remain broad even in the thermodynamic limit.

The probability distributions of the Fourier modes for the SP and CD models are broadly similar
except for the first mode $Q_1$. For the CD model, the distribution $P(Q_1)$ increases
monotonically with $Q_1$, while, for the SP model, it first increases and then decreases. This
difference reflects the fact that $Q_1$ reaches the maximum value of ${1}/{\pi} \simeq 0.318$
more often in the CD model than the SP model (See Fig.~\ref{fig:qtime}). Correspondingly the
value of $Q_1^{*} = \langle Q_1 \rangle$ is higher for CD model (See Table.~\ref{tab:1}).

The modes $Q_1, Q_2, Q_3, \ldots$ are anticorrelated at short times. This is already evident
from Fig.~\ref{fig:qtime}, which shows that local maxima $Q_2, Q_3, \ldots$ usually occur when
$Q_1$ is small. A quantification of the degree of anticorrelation and its behaviour in time
will be discussed in Section~\ref{sec:dynamics} below.

\figFour

\subsection{Scaling properties}

Let us turn to a discussion of the scaling behaviour of $Q^*(m,L) = \langle Q \left( {2\pi
m}/{L} \right) \rangle$ as a function of mode number $m$ and system size $L$. As noted earlier,
Fig.~\ref{fig:QkkCD} shows that: (a) For fixed value of $k = {2\pi m}/{L}$ away from zero,
$\langle Q(k)\rangle\rightarrow 0$ as $L\rightarrow \infty$. (b) For a fixed (typically small)
value of $m$, the mean value $Q_m^*$ approaches a finite $m$ dependent value as $L\rightarrow
\infty$.  Thus, in Fig.~\ref{fig:QkkCD} the limit $k\rightarrow 0$ exposes the set of order
parameters we have been discussing. The limiting values (as $L\rightarrow \infty$) of $Q_m^{*}$
fall with increasing $m$. In order to connect (a) and (b), we postulate the scaling form :
\begin{equation}
	Q^*(m,L) \sim L^{-\phi} Y \left( \frac{m}{L} \right).
\end{equation}
The fact that $k = {2\pi m}/{L}$ is a good variable in the large $L$ limit is already
incorporated in writing the argument $y$ as ${m}/{L}$. Evidently the scaling function
$Y(y)\rightarrow Y_0$, a constant value, as $y \rightarrow \infty$ in order to be consistent
with (a) above; $Q^*$ falls to zero as $Y_0 L^{-\phi}$. Turning to (b), the small-$m$ end of
Fig.~\ref{fig:QkkCD} corresponds to $y << 1$ where we expect $Y(y) \sim y^{-\gamma}$, which
would imply an $L$ dependence of the form $m^{-\gamma} L^{\gamma - \phi}$ for $Q^*(m,L)$. Since
the data in Fig.~\ref{fig:QkkCD} suggests finite limiting values as $L\rightarrow \infty$ as
noted in (b), we conclude that $\gamma = \phi$ and the limiting values $Q_m^*$ fall as
$m^{-\phi}$. This is verified in Fig.~\ref{fig:scaling} which shows the data for the SP model
for both EW and KPZ surface dynamics, and for the CD model. As noted earlier, the static
properties of the CD model are identical for both the EW and KPZ cases.

The observed scaling properties of the average value $Q^*(m,L)\equiv \langle Q \left( {2\pi
m}/{L} \right) \rangle$ suggest that the full probability distributions $P \left( {2\pi m}/{L}
\right)$ in Fig.~\ref{fig:probability} may exhibit a scaling collapse as well.  This is indeed
the case (except for first few modes), as seen in Fig.~\ref{fig:probscaling} where we have
plotted the scaled distribution ${P(Q_m)}/{m^{\phi}}$ versus $Q_m m^{\phi}$.

We conclude by commenting on the values of the exponent $\phi$. The values used in the plots in
Fig.~\ref{fig:scaling}  and Fig.~\ref{fig:probscaling} are $\phi=2/3$ and $\phi=3/5$ (for the
SP model with EW and KPZ surface dynamics, respectively) and $\phi=3/4$ (for the CD model). An
analytic understanding of these values would be very desirable.

\section{Order parameter Dynamics} \label{sec:dynamics}

\figFive

We now turn to a discussion of the scaling properties of time dependent correlation functions
in the models under study.

\subsection{Single site autocorrelation: earlier results}

An earlier study~\cite{sakuntala} has shown that the single site occupancy autocorrelation
function $A(t,L)\equiv\langle \sigma_i(0)\sigma_i(t)\rangle_L$ exhibits the scaling form
\begin{equation}
	A(t,L)\approx f \left( \frac{ t }{ L^z } \right),
\end{equation}
for both SP and CD models, with $z=2$ and $3/2$ for surface evolutions obeying EW and KPZ
dynamics respectively. The scaling function $f(y)$ was found to show a cusp singularity as
$y\rightarrow 0$
\begin{equation}
	f(y) \approx  m_0^2 \left[ 1 - b y^{\beta} \right] \quad \text{for} \quad  y<<1,
\end{equation}
where $m_0$ is a measure of LRO. For the SP model, the exponent $\beta$ was found to be
approximately $0.22$ for EW, and $0.18$ for KPZ dynamics. For the CD model, the
value $\beta=1/4$ was derived analytically for EW dynamics, while the value $\beta \simeq
0.31$ was found numerically for KPZ case~\cite{sakuntala}.

\subsection{Order parameter set: auto-correlation and cross-correlation}

\figSix

Let us define time-dependent correlation functions involving the order parameter set
\begin{equation}
	C_{mn}(t)= \frac{\langle Q_m(0)Q_n(t)\rangle -\langle Q_m\rangle\langle Q_n\rangle}
	{\langle  Q_m\rangle\langle Q_n\rangle}.
\end{equation}
In Fig.~\ref{fig:corrscale}, we have plotted $C_{mn}$ against the scaled argument $y=t/L^z$. The
collapse, evident in the figure, provides clear evidence of scaling. Moreover, the
cross-correlation function $C_{12}(t)$ is negative, implying that the Fourier modes for $m=1$
and $m=2$ are anticorrelated. This feature is evident in Fig.~\ref{fig:qtime} as well, and has
its origins in the fact that the configurations that contribute most to the two modes involve
respectively one large particle cluster or two large clusters, and are thus mutually exclusive.

\subsection{Dynamical structure functions and intermittency}

\figSeven

\figEight

We now turn to the characterization of the dynamical structure function of the first Fourier
mode $Q_1$, whose variation in time is shown in Fig.~\ref{fig:qtime}. The question arises
whether the time series $Q_1(t)$ is self-similar, or whether it displays intermittency. In this
subsection, we study the dynamical structure factors, and an associated measure, namely the
flatness, to answer this question. For the CD models, we find a weak but definite divergence of
the flatness in the limit of small argument, implying that the series is weakly intermittent.

The structure functions are defined as moments of the distributions of temporal variations of
$Q_1$
\begin{equation}
	S_n(t,L)= {\langle \left[ Q_1(t) -Q_1(0) \right]^n \rangle}_L.
\end{equation}
In general, $S_n (t, L)$ grows as a power of $t$, but saturates at a value which depends on
system size $L$. Figure~\ref{fig:S2S4} shows that $S_n (L, t)$ is a scaling function of the
variable $t/L^z$ for both $n = 2$ and $4$, where $z$ is the dynamical exponent. For large $t$,
$Q(t)$ and $Q(0)$ are independent, and thus the saturation values of $S_2$ and $S_4$ are given
by the static quantities $2( \langle Q_1^2 \rangle - \langle Q_1 \rangle^2)$ and $2( \langle
Q_1^4 \rangle - 8 \langle Q_1^3 \rangle \langle Q_1 \rangle + 6 \langle Q_1^2 \rangle^2)$
respectively. These values can be determined accurately from steady state data and are shown as
dashed lines in Fig.~\ref{fig:S2S4}.

The property of intermittency has to do with the behaviour of $S_n (y)$ at small values of $y
\equiv t/L^z$, i.e., $S_n \sim y^{\beta_n}$. For self-similar series, the exponents $\beta_n$
themselves grow linearly with $n$, i.e., $\beta_n / n$ is a constant. By contrast, for an
intermittent time series, this property is no longer true. A good diagnostic is the flatness
$\kappa(t/L^z) = S_4 / S_2^2$. For a self-similar system, $\kappa(y)$ approaches a constant
value as $y \rightarrow 0$, while for an intermittent system, $\kappa(y)$ diverges as
$y\rightarrow 0$, due to the preponderance of sharp rises and falls of $Q_1$ over relatively
short times.  The flatness as a function of scaled variable $t/L^z$ for both SP and CD models
with EW and KPZ dynamics is plotted in Fig.~\ref{fig:flat}.

Let us first consider the CD models, for both the EW and the KPZ dynamics. The data in
Fig.~\ref{fig:S2S4} indicate $S_2(y) \sim  y^{\beta_2}$ and $S_4(y) \sim  y^{\beta_4}$ with
$\beta_2 = 0.709 \pm 0.002$ (EW) and $0.907 \pm 0.003$ (KPZ), and  $\beta_4 = 1.365 \pm 0.005$
(EW) and $1.74 \pm 0.01$ (KPZ).  This would imply $\kappa(y)$ diverging as $y^{-\gamma}$ with
$\gamma = 2 \beta_2 - \beta_4$, leading to $\gamma = 0.05 \pm 0.01$ (EW) and $\gamma = 0.07 \pm
0.02$ (KPZ). The exponent $\gamma$ can also be be extracted directly from the plot for the
flatness (Fig.~\ref{fig:flat}). For the CD models we obtain $\gamma = 0.040 \pm 0.001$ (EW) and
$0.049 \pm 0.001$ (KPZ). There is a weak but definite divergence. Hence we conclude that the
time series $Q_1(t)$ displays weak intermittency for the CD models, in both the EW and KPZ
cases. For the SP models in the intermediate $y$ regime, we estimate $\beta_2 = 1.639 \pm
0.007$ (EW) and $1.700 \pm 0.006$ (KPZ), and $\beta_4 = 3.21 \pm 0.02$ (EW) and $3.33 \pm 0.02$
(KPZ). Correspondingly we find $\gamma = 0.07 \pm 0.03$ for both EW and KPZ dynamics. At
smaller values of $y$, the curves seems to show a flattening, but we are not able to reach a
firm conclusion owing to problems with the precision of the data in this case.

\section{Conclusions} \label{sec:conclude}

In this paper, we have examined the characterization of order in fluctuation-dominated phase
ordering (FDPO), by studying the static and dynamic properties of the order parameter set for a
system of sliding hard-core particles driven by a 1D fluctuating surface. In addition to the
sliding particle (SP) model, we studied the CD model of surface dynamics which mimics the
properties of the SP model. The necessity for having a set of order parameters -- here the
long-wavelength Fourier modes of the particle density -- arises from the fact that the
macroscopic cluster of particles has a finite probability of breaking into a small number $m$
of clusters ($m = 1,2,3,\ldots$); the occurrence of these clusters is tracked by the $m$th
Fourier mode, $Q_m$.

The characteristic of FDPO is that fluctuations remain large, and do not damp down in the
thermodynamic limit. Correspondingly, the probability distribution $P(Q_m)$ approaches a broad
limiting form as $L \rightarrow \infty$. The mean value $Q^*(m, L)$ is found to be a scaling
function of mode number $m$ and system size $L$, as is the full probability distribution. The
temporal properties of FDPO are influenced by the existence of an $L$ dependent time scale
$\sim L^z$ where $z$ is the dynamical exponent of the driving interface ($z= 2$ for EW, and
$3/2$ for KPZ case). The dynamical properties are a function of $t/L^z$. The most interesting
property is that the time series for the primary order parameter, $Q_1 (t)$, displays
intermittency in the case of the CD models for both EW and KPZ evolutions for the driving
interface. The signature of this is a weak but definite divergence of the flatness, which
involves the ratio of the fourth and second moments of the variations of the order parameter.
It would be interesting to be able to spell out the conditions for intermittency to arise in
systems displaying FDPO, and phase ordering in general.

\subsection*{Acknowledgements}

We thank Sanjay Puri for bringing to our attention the work on Porod law violations in
disordered systems. The authors acknowledge HPC facility at IISERM for generous computational
time.

\end{document}